\documentclass[sn-mathphys-num]{sn-jnl}
\usepackage{color}
\usepackage[usenames,dvipsnames,svgnames,table]{xcolor}

\usepackage{mathtools}
\usepackage{mathrsfs}%
\usepackage{graphicx}
\usepackage{dcolumn}
\usepackage{array}
\usepackage{lipsum}
\usepackage{bm}
\usepackage{subfigure}
\usepackage{amsmath,amssymb,amsfonts}%
\usepackage{amsthm}%
\usepackage{amstext} 
\usepackage{multirow}
\usepackage{tabularx}
\usepackage{braket}
\graphicspath{{plots/}}
\usepackage{comment}
\usepackage[title]{appendix}%
\usepackage{textcomp}%
\usepackage{manyfoot}%
\usepackage{booktabs}%
\usepackage{algorithm}%
\usepackage{algorithmicx}%
\usepackage{algpseudocode}%
\usepackage{listings}%

\newcolumntype{C}{>{$}c<{$}}

\newcommand{\beq}{\begin{equation}}
\newcommand{\eeq}{\end{equation}}
\newcommand{\bea}{\begin{eqnarray}}
\newcommand{\eea}{\end{eqnarray}}
\newcommand{\beal}{\begin{align}}
\newcommand{\eeal}{\end{align}}

\newcommand{\rv}{{\bf r}}
\newcommand{\kv}{{\bf k}}
\newcommand{\qv}{{\bf q}}

\newcommand{\xv}{{\bf x}}

\begin{document}
\title{
Green's function perspective on the nonlinear density response of quantum many-body systems}

\author*[1]{Jan Vorberger}
\email{j.vorberger@hzdr.de}
\affil[1]{Helmholtz-Zentrum Dresden-Rossendorf (HZDR), 01328 Dresden, Germany}

\author[2,1]{Tobias Dornheim}

\affil[2]{Center for Advanced Systems Understanding (CASUS), 02826 G\"orlitz, Germany}

\author[3]{Maximilian P.~B\"ohme}
\affil[3]{Lawrence Livermore National Laboratory (LLNL), California 94550 Livermore, USA}

\author[2,1]{Zhandos A.~Moldabekov}

\author[4]{Panagiotis Tolias}
\affil[4]{KTH Royal Institute of Technology, 114 28 Stockholm, Sweden}



\abstract{
 We derive equations of motion for higher order density response functions using the theory of thermodynamic Green's functions. We also derive expressions for the higher order generalized dielectric functions and polarization functions. Moreover, we relate higher order response functions and higher order collision integrals within the Martin-Schwinger hierarchy. We expect our results to be highly relevant to the study of a variety of quantum many-body systems such as matter under extreme temperatures, densities, and pressures.
}

\keywords{warm dense matter, density response, nonlinear response, Green's function, higher order correlations}

\maketitle

\section{Introduction}

The theory of correlations and fluctuations in non-ideal quantum Coulomb systems is well developed~\cite{kremp_book}. However, traditionally and by necessity or difficulty, mostly linear phenomena have been investigated and at most effective correlations between pairs of particles have been explicitly described. For calculations of thermodynamic properties, this is perfectly sufficient as all quantities can be determined purely on the basis of pair correlations. 

All higher order correlations are usually subsumed in local field corrections (LFC), effective fields, screened potentials, self energies and similar quantities~\cite{kremp_book}. However, dynamic correlations between three or four particles, as they are contained in LFCs or in the higher order collision integrals, also appear explicitly in the theory of density fluctuations. Here, the linear response functions can be computed using ideal response functions and LFCs. These LFCs are made up of quadratic, cubic, quartic etc. response functions. In addition, these response functions determine the quadratic and cubic density response as well as further orders. The nonlinear response functions describe the density fluctuations once the excitation is not infinitesimally small anymore, the nonlinear structure factors, and very many effects like higher harmonics generation, mode coupling and other interactions of density waves in matter~\cite{PhysRevE.54.3518,dornheim_prl_20,Dornheim_response_review_2023,Dornheim_imaginary_nonlinear_2021,Dornheim_nonlinear_2022}.

Here, we make explicit the equations of motion for the nonlinear density response functions. We show their relation to the nonlinear polarization functions and the nonlinear generalized dielectric response functions. We provide the connections between s-particle Green's \& correlation functions and the nonlinear response functions. We highlight problems in the theory that prevent us from computing certain contributions at the present time. 

Nonlinear fluctuations in (high temperature) plasmas have been investigated for a long time~\cite{Sitenko}. With the development of experimental warm dense matter physics and the corresponding diagnostics~\cite{siegfried_review,falk_wdm}, the investigation of density fluctuations is now fashionable again~\cite{Dornheim_response_review_2023,bonitz_pop_20}. In particular, high energy, high brilliance x-ray and optical lasers allow to study nonlinear phenomena~\cite{Fuchs_2015,Kettle_2021}.

Going beyond the linear response, sum rules and dissipation-fluctuation theorems have been established for the quadratic response~
\cite{Sitenko,PhysRevA.11.2147,PhysRevA.17.390,golden_pra_85,PhysRevE.54.3518,Kalman,CENNI1988279}.
First attempts were also made to improve quadratic responses with the help of LFCs~
\cite{Rennert,Paasch}. The general structure of the nonlinear density response matrix to monochromatic perturbations, for an arbitrary nonlinear order and an arbitrary harmonic, has been recently elucidated by Tolias {\em et al.}~\cite{Tolias_2023}. The same authors proved that the diagonal ideal nonlinear density response of arbitrary order can be expressed as the weighted sum of the ideal linear density responses evaluated at all multiple harmonics~\cite{Tolias_2023}, by building on pioneering work by Mikhailov~\cite{PhysRevLett.105.097401,Mikhailov_Annalen,Mikhailov_PRL}. An interesting application of nonlinear density fluctuations is given by the theoretical description of the stopping power~\cite{PhysRevB.37.9268,PhysRevB.56.15654,PhysRevB.52.13883,PhysRevB.59.10145}.

Quite recently, another dimension in the theoretical description of the correlations of density fluctuations has been opened with truly first principle descriptions of correlated, quantum warm dense matter. Such path integral Monte Carlo (PIMC) simulations using high performance computing facilities allow an approximation-free view at linear response quantities of the electron gas~
\cite{dornheim_prl_18,dynamic_folgepaper,hamann_prb_20,hamann_cpp_20} and also at real materials like hydrogen and beryllium~\cite{Boehme_snap_2022,Dornheim_H_response_2024,dornheim_H_LFC_2024}. Further, due to the non-perturbative nature of PIMC simulations, nonlinear density fluctuations and their correlations can be studied in various ways~\cite{Dornheim_PRL_2020,Dornheim_imaginary_nonlinear_2021,Dornheim_ferro_nonlinear_2021,Dornheim_corr_nonlinear_2021,dornheim2021density,Dornheim_nonlinear_2022,Tolias_2023}. These unambiguous results can then be used to inform xc-functionals and xc-kernels as they are needed as input quantities for density functional theory (DFT) calculations and time-dependent DFT calculations~\cite{Moldabekov_resp_rel_2021,Moldabekov_accurate_2022,Moldabekov_mix_2023,Moldabekov_kernel_2023}

In Section~\ref{sec_2}, we derive the equations of motion for the nonlinear density response functions. Section~\ref{sec_3} contains some elaborations concerning higher order generalized dielectric functions and polarization functions. Section~\ref{sec_4} reveals the connections between higher order Green's functions and higher order response functions as well as basic relations for the structure factors together with some considerations within the Martin-Schwinger hierarchy.

\section{Equations of motion for nonlinear density response functions\label{sec_2}}

The induced density $\delta n$ of a one-component quantum system under a small external perturbation $\delta U$ is usually taken to be~\cite{kremp_book}
\begin{equation}
\delta n(1)=\int d2\; L^{\mathrm{R}}(12)\delta U(2)\,,
\label{deltan}
\end{equation}
where $L^{\mathrm{R}}(12)=\mathrm{H}(t_1-t_2)\left[L^>(12)-L^<(12)\right]$ is the retarded version of the density-density fluctuation-correlation function with the greater and lesser components defined by $iL^>(12)=\langle\delta \rho(1) \delta \rho(2)\rangle$ and $iL^<(12)=\langle\delta \rho(2) \delta \rho(1)\rangle$, where $\mathrm{H}(\cdot)$ is the Heaviside step function. Note that $\hbar=1$ is used for simplicity everywhere. The operator of density fluctuations is $\delta\rho(1)=\psi^{\dagger}(1)\psi(1)-\langle\psi^{\dagger}(1)\psi(1)\rangle$, since $\rho(1)=\psi^{\dagger}(1)\psi(1)$, where $\psi$ and $\psi^{\dagger}$ stand for the standard field operators which automatically account for the proper quantum statistics and for which the shorthand notation of $1=\{\rv_1,t_1,\sigma_1\}$ has been employed for the space-spin coordinates.

The induced density is also given by the fluctuations of the one-particle correlation function $g^<$
\beq
\delta n(1)=-i\delta g^<(1,1^+,U)\,,
\eeq
where $1^+$ denotes an infinitesimally later time $t_1^+=t_1+0^+$.
Thus, Eq.~(\ref{deltan}) is a direct consequence of the definition of the response function $L$ in terms of the (external potential dependent) one-particle Green's function $g$~\cite{kremp_book}
\beq
L(12)=\pm i\frac{\delta g(11)}{\delta U(22)}
\label{def_L12}
\eeq
in an expansion around a small (vanishing) external potential $\delta U$ (the upper sign is for Bose statistics, the lower sign is for Fermi statistics). Relying on the smallness of the external perturbation, but taking into account terms beyond first order, one can immediately write the overall series expansion that now includes nonlinear responses as well
\bea
\delta n(1)&=&\int d2\; (\pm i) \frac{\delta g(11)}{\delta U(22)}\delta U(22)
+\frac{1}{2}\int d2d3\; (\pm i)^2 \frac{\delta^2 g(11)}{\delta U(22)\delta U(33)}\delta U(22)\delta U(33)\nonumber\\
&&+\frac{1}{6}\int d2d3d4\; (\pm i)^3 \frac{\delta^3 g(11)}{\delta U(22)\delta U(33)\delta U(44)}\delta U(22)\delta U(33)\delta U(44)+\ldots\nonumber\\
&&+\frac{1}{l!}\int \prod_{j=2}^{l+1}dj\; (\pm i)^l \frac{\delta^l g(11)}{\prod_{j=2}^{l+1}\delta U(jj)}\prod_{j=2}^{l+1}\delta U(jj)+\ldots\,.
\label{expansion}
\eea
This allows to read the definitions of higher order response functions off the Taylor expansion of the induced density. In particular, for the quadratic response $Y$, the cubic response $Z$ and the general $l-$order response $\mathcal{W}^{(l)}$, we obtain
\bea
Y(123)&=&(\pm i)^2 \frac{\delta^2 g(11)}{\delta U(22)\delta U(33)}\,,\\
Z(1234)&=&(\pm i)^3 \frac{\delta^3 g(11)}{\delta U(22)\delta U(33)\delta U(44)}\,,\\
\mathcal{W}^{(l)}(123...l+1)&=&(\pm i)^l \frac{\delta^l g(11)}{\prod_{j=2}^{l+1}\delta U(jj)}\,.
\eea
One might write these down as recursion formulas, in the following manner
\bea
Y(123)&=&(\pm i) \frac{\delta L(12)}{\delta U(33)}\,,\label{def_Y123}\\
Z(1234)&=&(\pm i) \frac{\delta Y(123)}{\delta U(44)}\,,\label{def_Z1234}\\
\mathcal{W}^{(l)}(123...l+1)&=&(\pm i) \frac{\delta \mathcal{W}^{(l-1)}(123...l)}{\delta U[(l+1)(l+1)]}\,.
\eea
These response functions are special cases of the more general higher order functions
\bea
L(12,1'2')&=&(\pm i) \frac{\delta g(11')}{\delta U(2'2)}\,,\label{def_L1212}\\
Y(123,1'2'3')&=&(\pm i)^2\frac{\delta^2g(11')}{\delta U(3'3)\delta U(2'2)}=(\pm i) \frac{\delta L(12,1'2')}{\delta U(3'3)}\,,\label{def_Y123123}\\
Z(1234,1'2'3'4')&=&(\pm i)^3 \frac{\delta^3 g(11')}{\delta U(4'4)\delta U(3'3)\delta U(2'2)} \\
&=&(\pm i) \frac{\delta Y(123,1'2'3')}{\delta U(4'4)}\,,\label{def_Z12341234}\\
\mathcal{W}^{(l)}(123...l+1,1'2'3'...l+1')&=&(\pm i)^l \frac{\delta^l g(11')}{\prod_{j=2}^{l+1}\delta U(j'j)}\\
&=&(\pm i) \frac{\delta \mathcal{W}^{(l-1)}(123...l,1'2'3'...l')}{\delta U[(l+1)'(l+1)]}\,.
\eea
It is evident that $L(12,1'2')$ may be reduced to $L(12)$ via $L(12)=L(12,1'2')\delta(11')\delta(22')$. The high-order function $L(12,1'2')$ is also closely related to the two-particle Green's function $g_2(12,1'2')$ and thus contains the correlated dynamics of pairs of particles (holes) in a many body system. Similarly, the next-in-order functions $Y$ and $Z$ are related to three- and four-particle Green's functions and describe complexes of three and four particles (holes), respectively. These expressions can be used to obtain equations of motion for all higher order functions.

\subsection{Linear response}

In any case, the equation of motion for the linear response function can be obtained from the integral form of the Dyson equation for the one particle Green's function by functional differentiation with respect to the external potential $U$, see Ref.~\cite{kremp_book} [page 120, Eq.~(4.16)]
\begin{equation}
L(12,1'2')=L_0(12,1'2')
\mp i \int d3 d4 d5 d6\; L_0(13,1'4)\frac{\delta \Sigma(43)}{\delta g(56)}L(52;62')\,,
\label{eoml}
\end{equation}
where we have the generalized four-point response function $L(12,1'2')$ and its free (ideal) part
\beq
L_0(12,1'2')=\pm i g(12')g(21')\,.
\eeq
The integral term captures all exchange and correlation contributions, all higher order correlations, all multi-particle interactions and complexes as they influence the density fluctuations described by $L$. 

The generalized local field correction $\theta(46,35)$ is given by the functional derivative of the self energy with respect to the one particle Green's function~\cite{kremp_book}
\begin{align}
\theta(46,35)=\frac{\delta \Sigma(43)}{\delta g(56)}
=&\pm i \delta(43)\delta(56)V(46)+\bar{\theta}(46,35)\\
=&\pm i \delta(43)\delta(56)V(46)+\frac{\delta\bar{\Sigma}(43)}{\delta g(56)}\,,
\label{lfc_lin}
\end{align}
Here, there is a first term containing just the Coulomb potential $V$ which gives rise to the mean field (RPA) term and a second term featuring the functional derivative of the screened self energy $\bar{\Sigma}=\Sigma-\Sigma^H$ (the Hartree term is subtracted), which contains all the higher order collisions, correlations and exchange, represents the LFC proper. 

Upon specification of the arguments as needed for the induced density as in Eq.~(\ref{def_L12}), the well-known bubble diagrams pop out like
\beq
L_0(12)=\pm i g(12)g(21)\;.
\eeq
Thus, the most general expression for the linear density response $L$ is
\bea
L(12)=L_0(12)\mp i \int d3 d4 d5 d6\; L_0(13,14)\theta(46,35)L(52;62)\,,\nonumber
\eea
where all higher order correlations are described by the generalized LFC $\theta(46,35)$. This equation still requires a three-point function $L(52;62)$ in the integral term on the right hand side and is thus not closed in this form. Only a simplification of the complicated general LFC  $\theta(46;35)$ into an, e.g., static limit, allows a similar mathematical structure as encountered in the RPA approximation.

In the RPA, which implies dropping the second term and therefore all LFC in Eq.~(\ref{lfc_lin}), we have
\beq
L^{\mathrm{RPA}}(12)=L_0(12)+ \int d3 d5\; L_0(13)V(35)L^{\mathrm{RPA}}(52)\,.
\eeq
This equation has the advantageous convolution structure which lends itself to Fourier transform
\beq
L^{\mathrm{RPA}}(\xv)=\frac{L_0(\xv)}{1-V(\kv)L_0(\xv)}\,.
\eeq
Here and everywhere, $\xv\rightarrow\left\{\kv,\omega\right\}$, $V(\kv)=V(k)$ and the spin variables have been dropped. It can be seen that such a structure holds for RPA as well as for the case of static LFCs. For completeness, we give the expression for the ideal linear density response function (sometimes called the Lindhard function)~\cite{kremp_book}
\beq
L_0(\kv,\omega)=
  \int\!\frac{d\qv}{(2\pi)^3}
  \frac{f_0(\qv)-f_0(\qv+\kv)}
  {\hbar\omega + i\epsilon - E_{0}(\qv + \kv) - E_{0}(\qv)}\,,
\eeq
with Fermi- or Bose functions $f_0$ and one-particle energies $E_0$.

\subsection{Quadratic response\label{quad_resp}}

The generalized six-point quadratic density response function follows directly without other external input from Eq.~(\ref{def_Y123123}) and Eq.~(\ref{eoml})

\bea
Y(123,1'2'3')&=&
-g(13')g(32')g(21')-g(12')g(23')g(31')\nonumber\\
&&\pm i D(13,2'3')g(21')\pm i g(12')D(23,1'3')\nonumber\\
&&+\int d4 d5 d6 d7\;\Big\{
\pm i g(13')g(35)g(41')\theta(57,46)L(62,72')\nonumber\\
&&\qquad\qquad\qquad\quad\;
\pm i g(15)g(43')g(31')\theta(57,46)L(62,72')\nonumber\\
&&\qquad\qquad\qquad\quad\; + D(13,53')g(41')\theta(57,46)L(62,72')\nonumber\\
&&\qquad\qquad\qquad\quad\; 
+ g(15)D(43,1'3')\theta(57,46)L(62,72')\nonumber\\
&&\qquad\qquad\qquad\quad\; \pm i \int d8 d9\; g(15)g(41')\theta(579,468)L(83,93')L(62,72')\nonumber\\
&&\qquad\qquad\qquad\quad\;\pm i g(15)g(41')\theta(57,46)Y(623,72'3')\Big\}\,.
\eea

Here, we have defined and used
\bea
D(12,1'2')&=&\int d3 d4 d5 d6\; g(14)g(31')\frac{\delta\Sigma(43)}{\delta g(56)}L(52;62')\,,\nonumber\\
\theta(57,46)&=&\frac{\delta\Sigma(54)}{\delta g(67)}\,,\nonumber\\
\theta(579,468)&=&\frac{\delta\theta(57,46)}{\delta g(89)}\,.
\eea
We might again define an ideal quadratic response function
\beq
Y_0(123,1'2'3')=-g(13')g(32')g(21')-g(12')g(23')g(31')\,,
\eeq
which then allows to rewrite the total quadratic response as
\begin{align}
Y(123,1'2'3')&=Y_0(123,1'2'3') 
\mp i \int d4 d5 d6 d7 \; Y_0(124,1'2'5)\theta(57,46)L(63,73')\nonumber\\
&\quad\mp i \int d4 d5 d6 d7\;Y_0(143,1'53')\theta(57,46)L(62,72')\nonumber\\
&\quad- \int d4 d5 d6 d7 d\bar{4} d\bar{5} d\bar{6} d\bar{7}\;Y_0(14\bar{4},1'5\bar{5})\theta(\bar{5}\bar{7},\bar{4}\bar{6})L(\bar{6}3,\bar{7}3')
\theta(57,46)L(62,72')\label{eomy}\\
&\quad+ \int d4 d5 d6 d7 d8 d9 \; L_0(14,1'5)\theta(579,468)L(83,93')L(62,72')\nonumber\\
&\quad\mp i \int d4 d5 d6 d7\;L_0(14,1'5)\theta(57,46)Y(623,72'3')\,.\nonumber
\end{align}
We note a typical coupling of the quadratic response to the linear response via linear LFCs in the first three integral terms. Then, we have a contribution linking linear response and quadratic LFCs and finally the last term linking the ideal linear response to the full quadratic one establishing the screening contributions.

\begin{figure}[th]
\begin{center}
\includegraphics[clip=true,width=0.65\textwidth]{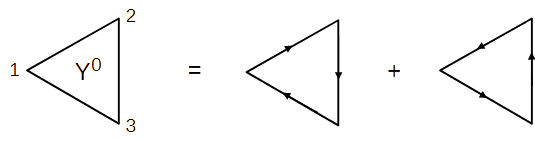}
\end{center}
\caption{\label{fig_y0}Feynman graphs of the ideal quadratic response function.}
\end{figure}

If we neglect all higher order terms beyond the pure Coulomb potential in the LFC type contributions $\theta(57,46)$ and completely neglect six-point functions like $\theta(579,486)$, the quadratic response function for the density response (\ref{def_Y123}) within the RPA can be obtained
\bea
Y^{\mathrm{RPA}}(123)&&=Y_0(123)
+\int d4 d5\; Y_0(124)V(45)L^{\mathrm{RPA}}(53)\nonumber\\
&&+\int d4 d5\; Y_0(143)V(45)L^{\mathrm{RPA}}(52)\nonumber\\
&&+\int d4 d5 d\bar{4} d\bar{5}\; Y_0(14\bar{4})V(45)L^{\mathrm{RPA}}(52)V(\bar{4}\bar{5})L^{\mathrm{RPA}}(\bar{5}3)\nonumber\\
&&+\int d4 d5\; L_0(14)V(45)Y^{\mathrm{RPA}}(523)\,.
\eea
The entire equation of motion has convolution structure and basically chains linear bubble diagrams to all corners of ideal quadratic response bubble diagrams (which look like triangles, see Fig.~\ref{fig_y0}).

This is also the justification for introducing static, linear LFCs in all the expressions, as they obey the same rules for collapsing variables/integrals as the potentials, thus $V(45)\to V(45)\left[1-G(45)\right]$ is allowed. As the structure in RPA is clearly of the convolution type, an algebraic expression is obtained in momentum-frequency space. For homogeneous and isotropic systems in k-space, one obtains
\beq
Y^{\mathrm{RPA}}(\xv_1,\xv_2)=\frac{Y_0(\xv_1,\xv_2)}
{
\varepsilon(\xv_1)
\varepsilon(\xv_2)
\varepsilon(\xv_1+\xv_2)
}\,,
\eeq
where the $\xv$-dependence again stands for $\{\kv,\omega\}$ and the linear dielectric function $\varepsilon(\xv)=1-V(\kv)L_0(\xv)$ was introduced.

We are left to determine the ideal quadratic response function. Its Fourier transform  is given by
\bea
Y_0(\xv_1,\xv_2)&=&\int\frac{d\qv_1}{(2\pi)^4} \Big\{
g(\qv_1)g(\qv_1-\xv_2)g(\qv_1-\xv_1-\xv_2)\nonumber\\
&&\qquad\qquad+g(\qv_1)g(\qv_1-\xv_1)g(\qv_1-\xv_1-\xv_2)
\Big\}\,,
\eea
where the integration symbol stands for momentum and frequency integration. For specific uses, the arguments can be further restricted. For instance, the second order part of the induced density in momentum-frequency space is
\beq
\delta n^{(2)}(\qv)=\frac{1}{2}\int\frac{d\xv_2}{(2\pi)^4}Y(\qv-\xv_2,\xv_2)\delta U(\qv-\xv_2)\delta U(\xv_2)\,,
\eeq
which demands specification of the quadratic response $Y$ according to
\begin{align}
Y^{\mathrm{RPA}}(\qv-\xv_2,\xv_2)&=\frac{Y_0(\qv-\xv_2,\xv_2)}
{\varepsilon(\qv-\xv_2)\varepsilon(\xv_2)
\varepsilon(\qv)}\,,\\
Y_0(\qv-\xv_2,\xv_2)=&\int\frac{d\qv_1}{(2\pi)^4} 
\big\{g(\qv_1)g(\qv_1-\xv_2)g(\qv_1-\qv)\nonumber\\
&\qquad\qquad+g(\qv_1)g(\qv_1-\qv+\xv_2)g(\qv_1-\qv)
\big\}\,.
\end{align}
If the external perturbing potential is monochromatic at $\qv_0$ with amplitude $A$, the sole remaining contribution of the quadratic response function is at the second harmonic
\begin{align}
\delta n^{(2)}(\qv)=\frac{A^2}{2}Y(\qv-\qv_0,\qv_0)&\delta(\qv-2\qv_0)\,,\\
Y^{\mathrm{RPA}}(\qv-\qv_0,\qv_0)\delta(\qv-2\qv_0)&=
\frac{
Y_0(\qv_0,\qv_0)
}
{\varepsilon(\qv_0)\varepsilon(\qv_0)\varepsilon(2\qv_0)
}\delta(\qv-2\qv_0)\,,\\
Y_0(\qv-\qv_0,\qv_0)\delta(\qv-2\qv_0)&=2\int\frac{d\qv_1}{(2\pi)^4}g(\qv_1)g(\qv_1-\qv_0)g(\qv_1-2\qv_0)\delta(\qv-2\qv_0)\,.
\end{align}
This can be evaluated either directly or with the well-known recursion formula for the quadratic response first given by Mikhailov~\cite{Mikhailov_Annalen,Mikhailov_PRL, Tolias_2023}. In case the external harmonic perturbation is not given by an exponential but is purely real, e.g. cosine-like, there will be a symmetric term at the negative second harmonic such that every $\qv_0$ is replaced by $-\qv_0$~\cite{Tolias_2023}.

\subsection{Cubic response}

The procedure is now repeated for the cubic response function, combining Eq.~(\ref{def_Z12341234}) with Eq.~(\ref{eomy}). We first give the ideal cubic response
\bea
Z_0(1234,1'2'3'4')&=&
\mp ig(14')g(43')g(32')g(21')
\mp ig(13')g(34')g(42')g(21')\nonumber\\
&&\mp ig(13')g(32')g(24')g(41')
\mp ig(14')g(42')g(23')g(31')\nonumber\\
&&\mp ig(12')g(24')g(43')g(31')
\mp ig(12')g(23')g(34')g(41')\,.
\eea
A rather lengthy and involved calculation with plenty of book--keeping and variable--tracking then gives for the total cubic response
\bea
\lefteqn{Z(1234,1'2'3'4')=Z_0(1234,1'2'3'4')}&&\nonumber\\
&&\mp i\int d5 d6 d7 d8\; Z_0(1235,1'2'3'6)\theta(68,57)L(74,74')\nonumber\\
&&\mp i \int d5 d6 d7 d8 \;Z_0(1254,1'2'64')\theta(68,57)L(73,83')\nonumber\\
&&\mp i \int d5 d6 d7 d8 \;Z_0(1534,1'63'4')\theta(68,57)L(72,82')\nonumber\\
&&-\int d5 d6 d7 d8 d\bar{5} d\bar{6} d\bar{7} d\bar{8}\; Z_0(125\bar{5},1'2'6\bar{6})
\theta(\bar{6}\bar{8},\bar{5}\bar{7})L(\bar{7}4,\bar{8}4')
\theta(68,57)L(73,83')\nonumber\\
&&-\int d5 d6 d7 d8 d\bar{5} d\bar{6} d\bar{7} d\bar{8}\; Z_0(153\bar{5},1'63'\bar{6})
\theta(\bar{6}\bar{8},\bar{5}\bar{7})L(\bar{7}4,\bar{8}4')
\theta(68,57)L(72,82')\nonumber\\
&&-\int d5 d6 d7 d8 d\bar{5} d\bar{6} d\bar{7} d\bar{8}\; Z_0(15\bar{5}5,1'6\bar{6}4')
\theta(\bar{6}\bar{8},\bar{5}\bar{7})L(\bar{7}3,\bar{8}3')
\theta(68,57)L(72,82')\nonumber\\
&&\pm i\int d5 d6 d7 d8 d\bar{5} d\bar{6} d\bar{7} d\bar{8}
d\bar{\bar{5}} d\bar{\bar{6}} d\bar{\bar{7}} d\bar{\bar{8}}\; Z_0(15\bar{5}\bar{\bar{5}},1'6\bar{6}\bar{\bar{6}})
\theta(\bar{\bar{6}}\bar{\bar{8}},\bar{\bar{5}}\bar{\bar{7}})
L(\bar{\bar{7}}4,\bar{\bar{8}}4')
\theta(\bar{6}\bar{8},\bar{5}\bar{7})L(\bar{7}3,\bar{8}3')
\theta(68,57)L(72,82')\nonumber\\
&&\mp i\int d5 d6 d7 d8 \;Y_0(125,1'2'6)\theta(68,57)Y(734,83'4')\nonumber\\
&&\mp i\int d5 d6 d7 d8 \;Y_0(153,1'63')\theta(68,57)Y(724,82'4')\nonumber\\
&&\mp i\int d5 d6 d7 d8 \;Y_0(154,1'64')\theta(68,57)Y(723,82'3')\nonumber\\
&&-\int d5 d6 d7 d8 d\bar{5} d\bar{6} d\bar{7} d\bar{8}\;
Y_0(15\bar{5},1'6\bar{6})\theta(\bar{6}\bar{8},\bar{5}\bar{7})L(\bar{7}4,\bar{8}4')
\theta(68,57)Y(723,82'3')\nonumber\\
&&-\int d5 d6 d7 d8 d\bar{5} d\bar{6} d\bar{7}d\bar{8}\;
Y_0(15\bar{5},1'6\bar{6})
\theta(\bar{6}\bar{8},\bar{5}\bar{7})Y(\bar{7}34,\bar{8}3'4')
\theta(68,57)L(72,82')\nonumber\\
&&-\int d5 d6 d7 d8 d\bar{5} d\bar{6} d\bar{7} d\bar{8}\;
Y_0(15\bar{5},1'6\bar{6})
\theta(\bar{6}\bar{8},\bar{5}\bar{7})L(\bar{7}3,\bar{8}3')
\theta(68,57)Y(724,82'4')\nonumber\\
&&-\int d5 d6 d7 d8 d\bar{5} d\bar{6} \; L_0(15,1'6)\theta(68\bar{6},57\bar{5})Y(\bar{5}34,\bar{6}3'4')L(72,82')\nonumber\\
&&-\int d5 d6 d7 d8 d\bar{5} d\bar{6} \; L_0(15,1'6)\theta(68\bar{6},57\bar{5})Y(724,82'4')L(\bar{5}3,\bar{6}3')\nonumber\\
&&-\int d5 d6 d7 d8 d\bar{5} d\bar{6} \; L_0(15,1'6)\theta(68\bar{6},57\bar{5})Y(723,82'3')L(\bar{6}4,\bar{5}4')\nonumber\\
&&\mp i \int d5 d6 d7 d8 d\bar{5} d\bar{6} \; Y_0(125,1'2'6)\theta(68\bar{5},57\bar{6})L(\bar{6}4,\bar{5}4')L(73,83')\nonumber\\
&&\mp i \int d5 d6 d7 d8 d\bar{5} d\bar{6} \; Y_0(153,1'63')\theta(68\bar{5},57\bar{6})L(\bar{6}4,\bar{5}4')L(72,82')\nonumber\\
&&\mp i \int d5 d6 d7 d8 d\bar{5} d\bar{6} \; Y_0(154,1'64')\theta(68\bar{6},57\bar{5})L(\bar{5}3,\bar{6}3')L(72,82')\nonumber\\
&&- \int d5 d6 d7 d8 d\bar{5} d\bar{6} d\bar{7} d\bar{8} d\bar{\bar{5}} d\bar{\bar{6}} \; Y_0(15\bar{5},1'6\bar{6})\theta(68\bar{\bar{5}},57\bar{\bar{6}})L(\bar{\bar{6}}4,\bar{\bar{5}}4')
L(\bar{7}3,\bar{8}3')\theta(68,57)L(72,82')\nonumber\\
&&-  \int d5 d6 d7 d8 d\bar{5} d\bar{6} d\bar{7} d\bar{8} d\bar{\bar{5}} d\bar{\bar{6}}\;
Y_0(15\bar{5},1'6\bar{6})\theta(68\bar{\bar{5}},57\bar{\bar{6}})L(\bar{\bar{6}}4,\bar{\bar{5}}4')
L(\bar{7}3,\bar{8}3')\theta(\bar{6}\bar{8},\bar{5}\bar{7})L(72,82')
\nonumber\\
&&- \int d5 d6 d7 d8 d\bar{5} d\bar{6} d\bar{7} d\bar{8} d\bar{\bar{5}} d\bar{\bar{6}} \;
Y_0(15\bar{5},1'6\bar{6})\theta(\bar{6}\bar{8},\bar{5}\bar{7})L(\bar{7}4,\bar{8}4')
\theta(68\bar{\bar{6}},57\bar{\bar{5}})L(\bar{\bar{5}}3,\bar{\bar{6}}3')L(72,82')\nonumber\\
&&\mp i \int d5 d6 d7 d8 d\bar{5} d\bar{6} d\bar{\bar{5}} d\bar{\bar{6}}\;L_0(15,1'6)\theta(68\bar{\bar{6}}\bar{5},57\bar{\bar{5}}\bar{6})L(\bar{6}4,\bar{5}4')L(\bar{\bar{5}}3,\bar{\bar{6}}3')L(72,82')\nonumber\\
&&\mp i \int d5 d6 d7 d8 \;L_0(15,1'6)\theta(68,57)Z(7234,82'3'4')
\,.
\eea
This includes the cubic local field correction
\beq
\theta(3579,2468)=\frac{\delta\theta(357,246)}{\delta g(89)}\,.
\eeq
Again, we observe coupling of the cubic response to the linear and quadratic responses but also coupling to functions of higher order. In the first three terms, the ideal quadratic response function is stuffed with bubbles on one corner; then follow three combinations of $Z_0$ with two linear response functions and a single term featuring $Z_0$ and three $L$'s. Then, we have three terms of combinations of quadratic response functions and linear LFCs followed by three terms combining linear and quadratic response functions with linear LFCs. The next six terms feature quadratic LCFs and quadratic responses combined with linear responses. Finally, there are terms that combine linear and quadratic LFCs with linear and quadratic response functions. The last term closes the integral equation in $Z$.

The diagonalization as required for the density perturbation yields first for the ideal cubic response
\bea
Z_0(1234)&=&
\mp ig(12)g(23)g(34)g(41)
\mp ig(14)g(43)g(32)g(21)\nonumber\\
&&\mp ig(13)g(34)g(42)g(21)
\mp ig(13)g(32)g(24)g(41)\nonumber\\
&&\mp ig(14)g(42)g(23)g(31)
\mp ig(12)g(24)g(43)g(31)\,,
\eea
and subsequently for the RPA approximation 
\bea 
\lefteqn{Z^{\mathrm{RPA}}(1234)=Z_0(1234)
+\int d5 d6 \;L_0(15)V(56)Z^{\mathrm{RPA}}(6234)}&&\nonumber\\
&&+\int d5 d6\; Z_0(1235)V(56)L^{\mathrm{RPA}}(64)
+\int d5 d6 \;Z_0(1254)V(56)L^{\mathrm{RPA}}(63)\nonumber\\
&&+\int d5 d6 \;Z_0(1534)V(56)L^{\mathrm{RPA}}(62)\nonumber\\
&&+\int d5 d6 d\bar{5} d\bar{6}\; Z_0(125\bar{5})V(\bar{5}\bar{6})L^{\mathrm{RPA}}(\bar{6}4)
V(56)L^{\mathrm{RPA}}(63)\nonumber\\
&&+\int d5 d6 d\bar{5} d\bar{6}\; Z_0(153\bar{5})V(\bar{5}\bar{6})L^{\mathrm{RPA}}(\bar{6}4)
V(56)L^{\mathrm{RPA}}(62)\nonumber\\
&&+\int d5 d6 d\bar{5} d\bar{6}\; Z_0(15\bar{5}4)V(\bar{5}\bar{6})L^{\mathrm{RPA}}(\bar{6}3)
V(56)L^{\mathrm{RPA}}(62)\nonumber\\
&&+\int d5 d6 d\bar{5} d\bar{6}
d\bar{\bar{5}}  d\bar{\bar{6}}\; Z_0(15\bar{5}\bar{\bar{5}})
V(\bar{\bar{5}}\bar{\bar{6}})
L^{\mathrm{RPA}}(\bar{\bar{6}}4)
V(\bar{5}\bar{6})L^{\mathrm{RPA}}(\bar{6}3)
V(56)L^{\mathrm{RPA}}(62)\nonumber\\
&&+\int d5 d6\;Y_0(125)V(56)Y^{\mathrm{RPA}}(634)
+\int d5 d6\;Y_0(153)V(56)Y^{\mathrm{RPA}}(624)\nonumber\\
&&+\int d5 d6\;Y_0(154)V(56)Y^{\mathrm{RPA}}(623)\nonumber\\
&&+\int d5 d6 d\bar{5} d\bar{6}\;
Y_0(15\bar{5})V(\bar{5}\bar{6})L^{\mathrm{RPA}}(\bar{6}4)
V(56)Y^{\mathrm{RPA}}(623)\nonumber\\
&&+\int d5 d6 d\bar{5}  d\bar{6}\;
Y_0(15\bar{5})
V(\bar{5}\bar{6})Y^{\mathrm{RPA}}(\bar{6}34)
V(56)L^{\mathrm{RPA}}(62)\nonumber\\
&&+\int d5 d6 d\bar{5}  d\bar{6}\;
Y_0(15\bar{5})
V(\bar{5}\bar{6})L^{\mathrm{RPA}}(\bar{6}3)
V(56)Y^{\mathrm{RPA}}(624)\,.
\eea
\begin{figure*}[ht]
\includegraphics[width=\textwidth,clip=true]{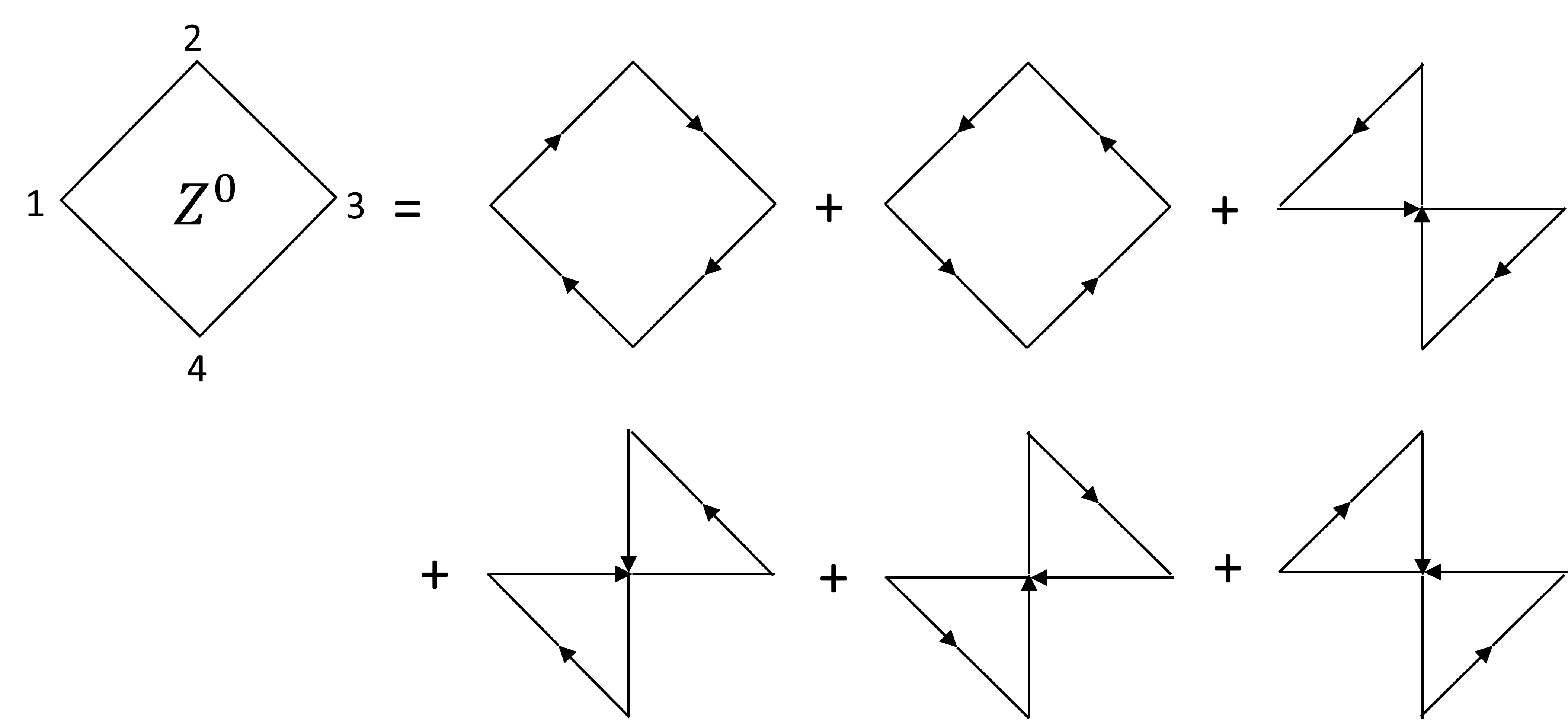}
\caption{\label{fig_z0} The Feynman graphs representing the terms comprising the ideal cubic density response function $Z^0(1234)$.}
\end{figure*}
There is one bubble in linear response, two bubbles in quadratic response, and six for the cubic response. A graphical representation of the topology of these terms is given in Fig.~\ref{fig_z0}. It is clear that these terms arise when considering all the different possibilities of connecting four points; the actual number of terms is given by the factorial of the order of the response.

A Fourier transform seems prudent again and the result is
\bea
\lefteqn{Z^{\mathrm{RPA}}(\xv_1,\xv_2,\xv_3)=}&&\nonumber\\
&&\frac
{Z_{0}(\xv_1,\xv_2,\xv_3)}
{
\left[1-V(\kv_1)L_0(\xv_1)\right]
\left[1-V(\kv_2)L_0(\xv_2)\right]
\left[1-V(\kv_3)L_0(\xv_3)\right]
\left[1-V(\kv_1+\kv_2+\kv_3)L_0(\xv_1+\xv_2+\xv_3)\right]
}
\nonumber\\
&+&
\frac{1}{\left[1-V(\kv_1+\kv_2+\kv_3)L_0(\xv_1+\xv_2+\xv_3)\right]}
\Big\{
Y_0(-\xv_1-\xv_2,\xv_3)V(\kv_1+\kv_2)Y(\xv_1,\xv_2)\\
&&\qquad\qquad\qquad\qquad\qquad\qquad+Y_0(\xv_1,-\xv_2-\xv_3)V(\kv_2+\kv_3)Y(\xv_2,\xv_3)\nonumber\\
&&\qquad\qquad\qquad\qquad\qquad\qquad+Y_0(-\xv_1-\xv_3,\xv_2)V(\kv_1+\kv_3)Y(\xv_1,\xv_3)\nonumber\\
&&\qquad\qquad\qquad\qquad\qquad\qquad+Y_0(\xv_1+\xv_2,\xv_3)V(\kv_1+\kv_2)Y(\xv_1,\xv_2)V(\kv_3)L(\xv_3)\nonumber\\
&&\qquad\qquad\qquad\qquad\qquad\qquad+Y_0(\xv_1,\xv_2+\xv_3)V(\kv_2+\kv_3)Y(\xv_2,\xv_3)V(\kv_1)L(\xv_1)\nonumber\\
&&\qquad\qquad\qquad\qquad\qquad\qquad+Y_0(\xv_1+\xv_3,\xv_2)V(\kv_1+\kv_3)Y(\xv_1,\xv_3)V(\kv_2)L(\xv_2)\Big\}\,.\nonumber
\eea
It is obvious that the structure is different to the expected one from the quadratic response. The first line is still familiar in its structure with the ideal cubic response in the numerator and a denominator formed by combining various linear dielectric functions. However, everything beyond the first line has no equal in the RPA expressions of the linear and quadratic response functions.

Fourier transforming the ideal cubic response gives
\bea
Z_0(\xv_1,\xv_2,\xv_3)&=&\int\frac{d\qv_1}{(2\pi)^4}g(\qv_1)g(\qv_1-\xv_1-\xv_2-\xv_3)
\nonumber\\
&&\times\Big\{g(\qv_1-\xv_3)g(\qv_1-\xv_2-\xv_3)\nonumber\\
&&\quad+g(\qv_1-\xv_3)g(\qv_1-\xv_1-\xv_3)\nonumber\\
&&\quad+g(\qv_1-\xv_2)g(\qv_1-\xv_2-\xv_3)\nonumber\\
&&\quad+g(\qv_1-\xv_2)g(\qv_1-\xv_1-\xv_2)\nonumber\\
&&\quad+g(\qv_1-\xv_1)g(\qv_1-\xv_1-\xv_3)\nonumber\\
&&\quad+g(\qv_1-\xv_1)g(\qv_1-\xv_1-\xv_2)\Big\}\,.
\eea
For the induced density due to the cubic response 
\begin{align}
\delta n^{(3)}(\qv)&=\frac{1}{6}\int\frac{d\xv_2}{(2\pi)^4}\frac{d\xv_3}{(2\pi)^4}Z(\qv-\xv_2-\xv_3,\xv_2,\xv_3)\delta U(\qv-\xv_2-\xv_3)\delta U(\xv_2)\delta U(\xv_3)\,,
\label{cub_dens_basic}
\end{align}
we need a particular case of the corresponding cubic response function in RPA 
{\small
\begin{align}
\MoveEqLeft Z^{\mathrm{RPA}}(\qv-\xv_2-\xv_3,\xv_2,\xv_3)=&\nonumber\\
&\frac
{Z_{0}(\qv-\xv_2-\xv_3,\xv_2,\xv_3)}
{
\left[1-V(\qv-\kv_2-\kv_3)L_0(\qv-\xv_2-\xv_3)\right]
\left[1-V(\kv_2)L_0(\xv_2)\right]
\left[1-V(\kv_3)L_0(\xv_3)\right]
\left[1-V(\qv)L_0(\qv)\right]
}
\nonumber\\
&+\frac{1}{\left[1-V(\qv)L_0(\qv)\right]}
\Bigg\{
\frac{Y_0(-\qv+\xv_3,\xv_3)V(\qv-\kv_3)Y_0(\qv-\xv_2-\xv_3,\xv_2)}
{\left[1-V(\qv-\kv_2-\kv_3)L_0(\qv-\xv_2-\xv_3)\right]\left[1-V(\kv_2)L_0(\xv_2)\right]
\left[1-V(\qv-\kv_3)L_0(\qv-\xv_3)\right]}
\nonumber\\
&\qquad\qquad\qquad\qquad
+\frac{Y_0(\qv-\xv_2-\xv_3,-\xv_2-\xv_3)V(\kv_2+\kv_3)Y_0(\xv_2,\xv_3)}
{\left[1-V(\kv_2)L_0(\xv_2)\right]\left[1-V(\kv_3)L_0(\xv_3)\right]
\left[1-V(\kv_2+\kv_3)L_0(\xv_2+\xv_3)\right]}
\nonumber\\
&\qquad\qquad\qquad\qquad
+\frac{Y_0(-\qv+\xv_2,\xv_2)V(\qv-\kv_2)Y_0(\qv-\xv_2-\xv_3,\xv_3)}
{\left[1-V(\qv-\kv_2-\kv_3)L_0(\qv-\xv_2-\xv_3)\right]\left[1-V(\kv_3)L_0(\xv_3)\right]
\left[1-V(\qv-\kv_2)L_0(\qv-\xv_2)\right]}\Bigg\}\\
&
+\frac{Y_0(\qv-\xv_3,\xv_3)V(\qv-\kv_3)Y_0(\qv-\xv_2-\xv_3,\xv_2)V(\kv_3)L_0(\xv_3)}
{\left[1-V(\qv-\kv_2-\kv_3)L_0(\qv-\xv_2-\xv_3)\right]
\left[1-V(\qv-\kv_3)L_0(\qv-\xv_3)\right]
\left[1-V(\kv_2)L_0(\xv_2)\right]\left[1-V(\kv_3)L_0(\xv_3)\right]}\nonumber\\
&
+\frac{Y_0(\qv-\xv_2-\xv_3,\xv_2+\xv_3)V(\kv_2+\kv_3)Y_0(\xv_2,\xv_3)V(\qv-\kv_2-\kv_3)L_0(\qv-\xv_2-\xv_3)}
{\left[1-V(\qv-\kv_2-\kv_3)L_0(\qv-\xv_2-\xv_3)\right]
\left[1-V(\kv_2+\kv_3)L_0(\xv_2+\xv_3)\right]
\left[1-V(\kv_2)L_0(\xv_2)\right]\left[1-V(\kv_3)L_0(\xv_3)\right]}
\nonumber\\
&
+\frac{Y_0(\qv-\xv_2,\xv_2)V(\qv-\kv_2)Y_0(\qv-\xv_2-\xv_3,\xv_3)V(\kv_2)L_0(\xv_2)}
{\left[1-V(\qv-\kv_2-\kv_3)L_0(\qv-\xv_2-\xv_3)\right]
\left[1-V(\qv-\kv_2)L_0(\qv-\xv_3)\right]
\left[1-V(\kv_2)L_0(\xv_2)\right]\left[1-V(\kv_3)L_0(\xv_3)\right]}
\,.\nonumber
\end{align}
}
Its ideal part is
\begin{align}
Z_0(\qv-\xv_2-\xv_3,\xv_2,\xv_3)&=\int\frac{d\qv_1}{(2\pi)^4}
g(\qv_1)g(\qv_1-\qv)\Big\{g(\qv_1-\xv_3)g(\qv_1-\xv_2-\xv_3)\nonumber\\
 &\qquad\qquad\qquad\qquad+g(\qv_1-\xv_3)g(\qv_1-\qv+\xv_2)\nonumber\\
 &\qquad\qquad\qquad\qquad+g(\qv_1-\xv_2)g(\qv_1-\xv_2-\xv_3)\nonumber\\
 &\qquad\qquad\qquad\qquad+g(\qv_1-\xv_2)g(\qv_1-\qv+\xv_3)\nonumber\\
 &\qquad\qquad\qquad\qquad+g(\qv_1-\qv+\xv_2+\xv_3)g(\qv_1-\qv+\xv_2)\nonumber\\
 &\qquad\qquad\qquad\qquad+g(\qv_1-\qv+\xv_2+\xv_3)g(\qv_1-\qv+\xv_3)\Big\}\,.
\end{align}
If the perturbing potential in Eq.~(\ref{cub_dens_basic}) is monochromatic at $\qv_0$ with amplitude $A$, there arise two distinct contributions from the cubic response function (assuming a complex exponential, see the argument for the density response at the second harmonic in section~\ref{quad_resp}):
\begin{align}
\delta n^{(3)}(\qv)=&\frac{A^3}{6}\Big\{
Z(\qv,\qv_0,-\qv_0)\delta(\qv-\qv_0)
+Z(\qv,-\qv_0,\qv_0)\delta(\qv-\qv_0)\nonumber\\
&+Z(\qv-2\qv_0,\qv_0,\qv_0)\delta(\qv-\qv_0)
+Z(\qv-2\qv_0,\qv_0,\qv_0)\delta(\qv-3\qv_0)\,.
\Big\}
\label{cubharm}
\end{align}
The cubic response function changes the response of the system at the first harmonic [first three adders of Eq.~(\ref{cubharm})] and also produces the contribution at the third harmonic [last adder of Eq.~(\ref{cubharm})]. This is the third-order realization of the more general behavior of any $l$-th order response function to change the density responses at all $l-2m>0$ harmonics and to introduce the leading order response at the $l$-th harmonic~\cite{Tolias_2023}.

\subsubsection{Cubic response at the third harmonic}

After executing the $\delta$-functions, the response function at the third harmonic is 
\bea
\lefteqn{
\chi^{(3,3)}(\qv_0)=\frac{\chi_0^{(3,3)}(\qv_0)}
{\left[1-V(\qv_0)\chi_0^{(1,1)}(\qv_0)\right]^3\left[1-V(3\qv_0)\chi_0^{(1,1)}(3\qv_0)\right]}}&&\\
&&+3\frac{Y_0(-2\qv_0,\qv_0)V(2\qv_0)\chi_0^{(2,2)}(\qv_0)}
{\left[1-V(\qv_0)\chi_0^{(1,1)}(\qv_0)\right]^2\left[1-V(2\qv_0)\chi_0^{(1,1)}(2\qv_0)\right]
\left[1-V(3\qv_0)\chi_0^{(1,1)}(3\qv_0)\right]}\nonumber\\
&&+3\frac{Y_0(2\qv_0,\qv_0)V(2\qv_0)\chi_0^{(2,2)}(\qv_0)V(\qv_0)\chi_0^{(1,1)}(\qv_0)}
{\left[1-V(\qv_0)\chi_0^{(1,1)}(\qv_0)\right]^3\left[1-V(2\qv_0)\chi_0^{(1,1)}(2\qv_0)\right]
\left[1-V(3\qv_0)\chi_0^{(1,1)}(3\qv_0)\right]}\nonumber
\eea
where it was used that $Z(\qv_0,\qv_0,\qv_0)=\chi^{(3,3)}(\qv_0)$, $Y(\qv_0,\qv_0)=\chi^{(2,2)}(\qv_0)$, and $L(\qv_0)=\chi^{(1,1)}(\qv_0)$. Also, it holds that $Y_0(-2\qv_0,\qv_0)=\chi_0^{(2,2)}(\qv_0)$. The general Fourier space connection of the $l$-order density response $\mathcal{W}^{(l)}$ with the harmonic density response $\chi^{(m,l)}$ for an arbitrary nonlinear order $l$ and an arbitrary harmonic $m$ has been elucidated in Ref.\cite{Tolias_2023}.

Finally, the ideal cubic response function consists of six identical terms
\begin{equation}
\chi_0^{(3,3)}(\qv_0)=Z_0(\qv_0,\qv_0,\qv_0)=6\int\frac{d\qv_1}{(2\pi)^4}g(\qv_1)g(\qv_1-\qv_0)g(\qv_1-2\qv_0)g(\qv_1-3\qv_0)\,.
\end{equation}
This expression is the same as given by Mikhailov and obeys the known recursion formula~\cite{Mikhailov_PRL,Tolias_2023}.

\subsubsection{Cubic response at the first harmonic}

For the cubic response function at the first harmonic, we obtain
\bea
\lefteqn{Z(\qv,\qv_0,-\qv_0)\delta(\qv-\qv_0)=
\frac{Z_0(\qv,\qv_0,-\qv_0)\delta(\qv-\qv_0)}
{
\left[1-V(\qv)L_0(\qv)\right]^2
\left[1-V(\qv_0)L_0(\qv_0)\right]^2
}}&&\nonumber\\
&+&\frac{\delta(\qv-\qv_0)}{\left[1-V(\qv)L_0(\qv)\right]}\Bigg\{
\frac{Y_0(-\qv-\qv_0,-\qv_0)V(\qv+\qv_0)Y_0(\qv,\qv_0)}
{\left[1-V(\qv)L_0(\qv)\right]\left[1-V(\qv_0)L_0(\qv_0)\right]
\left[1-V(\qv+\qv_0)L_0(\qv+\qv_0)\right]}
\nonumber\\
&&\qquad\qquad
+\frac{Y_0(\qv,0)V(0)Y_0(\qv_0,-\qv_0)}
{\left[1-V(\qv_0)L_0(\qv_0)\right]\left[1-V(-\qv_0)L_0(-\qv_0)\right]
\left[1-V(0)L_0(0)\right]}
\nonumber\\
&&\qquad\qquad
+\frac{Y_0(-\qv+\qv_0,\qv_0)V(\qv-\qv_0)Y_0(\qv,-\qv_0)}
{\left[1-V(\qv)L_0(\qv)\right]\left[1-V(-\qv_0)L_0(-\qv_0)\right]
\left[1-V(\qv-\qv_0)L_0(\qv-\qv_0)\right]}\Bigg\}
\,,
\\
\lefteqn{Z(\qv,-\qv_0,\qv_0)\delta(\qv-\qv_0)=
\frac{Z_0(\qv,-\qv_0,\qv_0)\delta(\qv-\qv_0)}
{
\left[1-V(\qv)L_0(\qv)\right]^2
\left[1-V(\qv_0)L_0(\qv_0)\right]^2
}}&&\nonumber\\
&+&\frac{\delta(\qv-\qv_0)}{\left[1-V(\qv)L_0(\qv)\right]}\Bigg\{
\frac{Y_0(-\qv+\qv_0,\qv_0)V(\qv-\qv_0)Y_0(\qv,-\qv_0)}
{\left[1-V(\qv)L_0(\qv)\right]\left[1-V(-\qv_0)L_0(-\qv_0)\right]
\left[1-V(\qv-\qv_0)L_0(\qv-\qv_0)\right]}
\nonumber\\
&&\qquad\qquad
+\frac{Y_0(\qv,0)V(0)Y_0(-\qv_0,\qv_0)}
{\left[1-V(-\qv_0)L_0(-\qv_0)\right]\left[1-V(\qv_0)L_0(\qv_0)\right]
\left[1-V(0)L_0(0)\right]}
\nonumber\\
&&\qquad\qquad
+\frac{Y_0(-\qv-\qv_0,-\qv_0)V(\qv+\qv_0)Y_0(\qv,\qv_0)}
{\left[1-V(\qv)L_0(\qv)\right]\left[1-V(\qv_0)L_0(\qv_0)\right]
\left[1-V(\qv+\qv_0)L_0(\qv+\qv_0)\right]}\Bigg\}
\,,
\\
\lefteqn{Z(\qv-2\qv_0,\qv_0,\qv_0)\delta(\qv-\qv_0)=}&&\nonumber\\
&&\frac{Z_0(\qv-2\qv_0,\qv_0,\qv_0)\delta(\qv-\qv_0)}
{
\left[1-V(\qv-2\qv_0)L_0(\qv-2\qv_0)\right]
\left[1-V(\qv_0)L_0(\qv_0)\right]^2
\left[1-V(\qv)L_0(\qv)\right]
}\nonumber\\
&+&\frac{\delta(\qv-\qv_0)}{\left[1-V(\qv)L_0(\qv)\right]}\nonumber\\
&&\times\Bigg\{
\frac{Y_0(-\qv+\qv_0,\qv_0)V(\qv-\qv_0)Y_0(\qv-2\qv_0,\qv_0)}
{\left[1-V(\qv-2\qv_0)L_0(\qv-2\qv_0)\right]\left[1-V(\qv_0)L_0(\qv_0)\right]
\left[1-V(\qv-\qv_0)L_0(\qv-\qv_0)\right]}
\nonumber\\
&&
\quad+\frac{Y_0(\qv-2\qv_0,0)V(2\qv_0)Y_0(\qv_0,\qv_0)}
{\left[1-V(\qv_0)L_0(\qv_0)\right]\left[1-V(\qv_0)L_0(\qv_0)\right]
\left[1-V(2\qv_0)L_0(2\qv_0)\right]}
\nonumber\\
&&
\quad+\frac{Y_0(-\qv+\qv_0,\qv_0)V(\qv-\qv_0)Y_0(\qv-2\qv_0,\qv_0)}
{\left[1-V(\qv-2\qv_0)L_0(\qv-2\qv_0)\right]\left[1-V(\qv_0)L_0(\qv_0)\right]
\left[1-V(\qv-\qv_0)L_0(\qv-\qv_0)\right]}\Bigg\}
\,,\\
\lefteqn{\chi^{(1,3)}(\qv_0)=Z(\qv_0,\qv_0,-\qv_0)+Z(\qv_0,-\qv_0,\qv_0)+Z(-\qv_0,\qv_0,\qv_0)}
\eea
Immediately, terms containing $V(0)\to\infty$ can be observed in some numerators and denominators of the same summands. In the denominator, such factors are combined with linear response functions at zero argument. In the numerators, quadratic response functions with one argument vanishing appear. We thus suggest that such terms need to cancel and the final result be finite since the effect of the cubic response at the first harmonic is proven in experiment and via numerical simulations.

The ideal cubic response at the first harmonic is given by
\begin{multline}
\chi_0^{(1,3)}(\qv_0)=Z_0(\qv_0,\qv_0,-\qv_0)+Z_0(\qv_0,-\qv_0,\qv_0)
+Z_0(-\qv_0,\qv_0,\qv_0)\\
=6\int\frac{d\qv_1}{(2\pi)^4}
g(\qv_1)g(\qv_1-\qv_0)\big\{
g(\qv_1+\qv_0)g(\qv_1)
+g(\qv_1-\qv_0)g(\qv_1)
+g(\qv_1-\qv_0)g(\qv_1-2\qv_0)
\big\}
\end{multline}
Up to now, nobody has been able to derive an analytical expression, a recursion formula or even a numerically solvable formula for the cubic response at the first harmonic. There appear infinities in the evaluation that so far have been intractable. Further, such infinities are expected for any off-diagonal element of the harmonic ideal density response matrix $\mathcal{X}_0=\{\chi_0^{(m,l)}\}$, i.e., they always emerge when $l$-order nonlinearities influence lower $m$-order harmonics~\cite{Tolias_2023}.

\begin{figure*}[th]
\begin{center}
\includegraphics[clip=true,width=0.8\textwidth]
{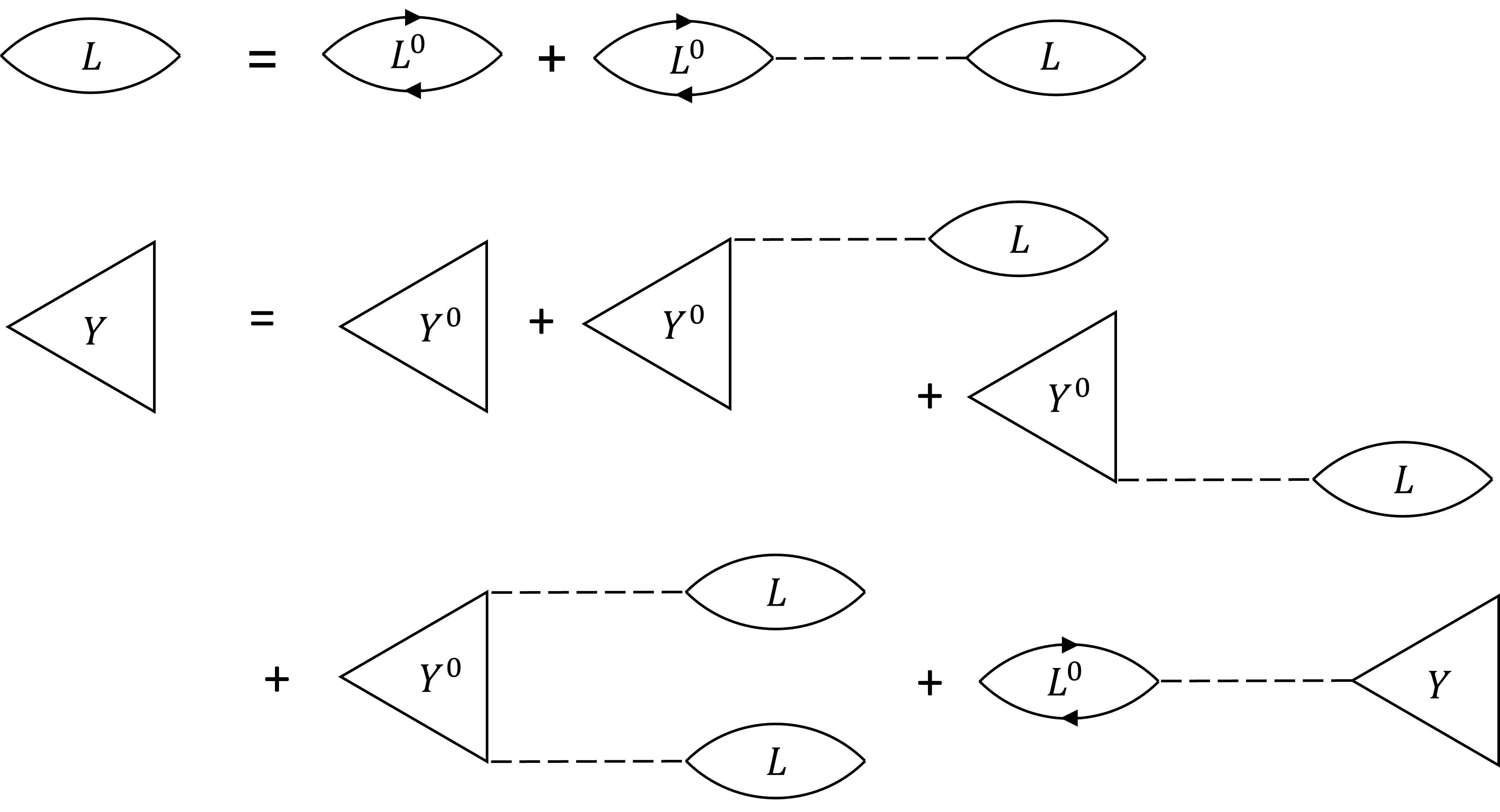}
\includegraphics[clip=true,width=0.8\textwidth]
{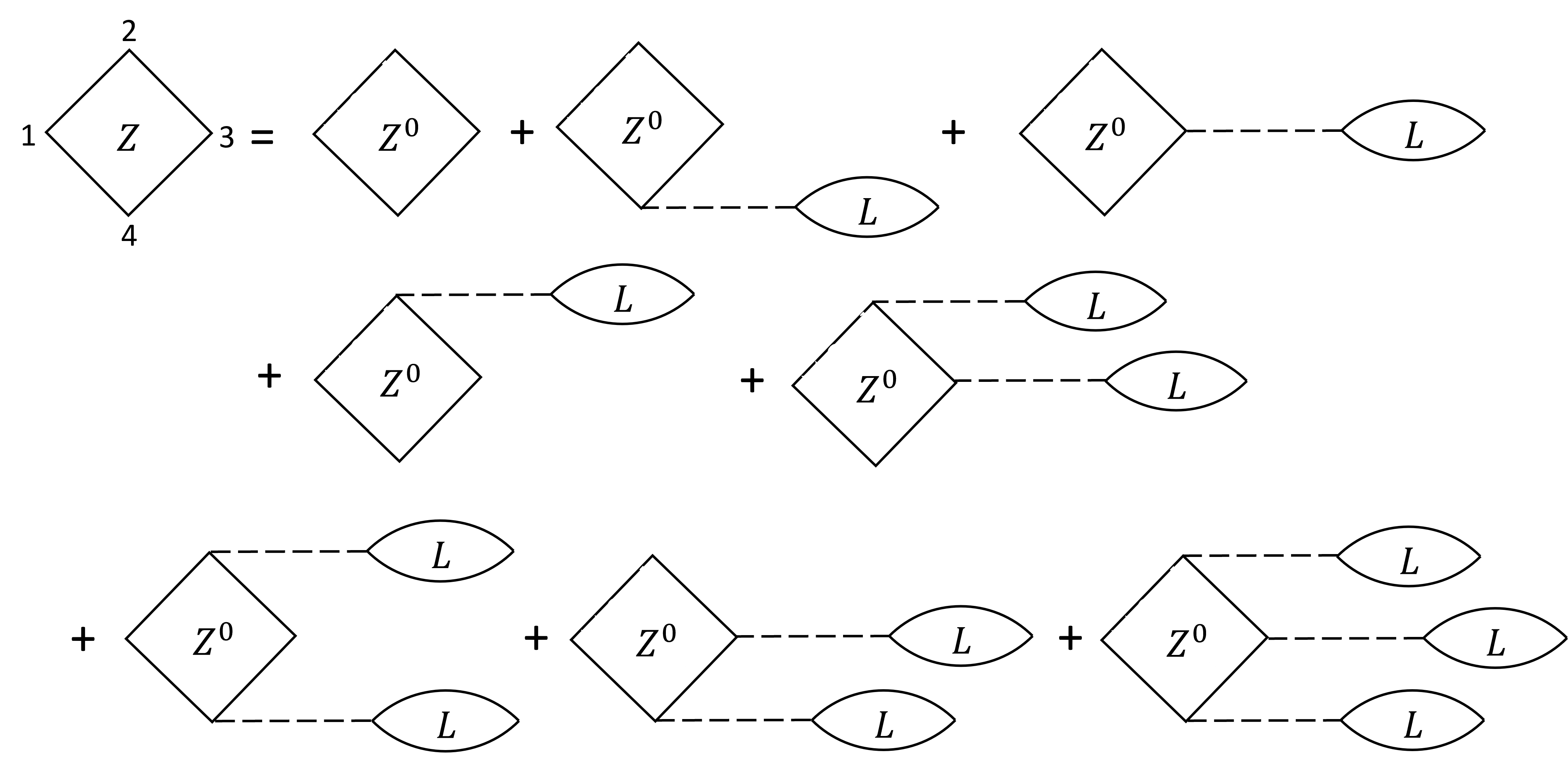}
\includegraphics[clip=true,width=0.8\textwidth]
{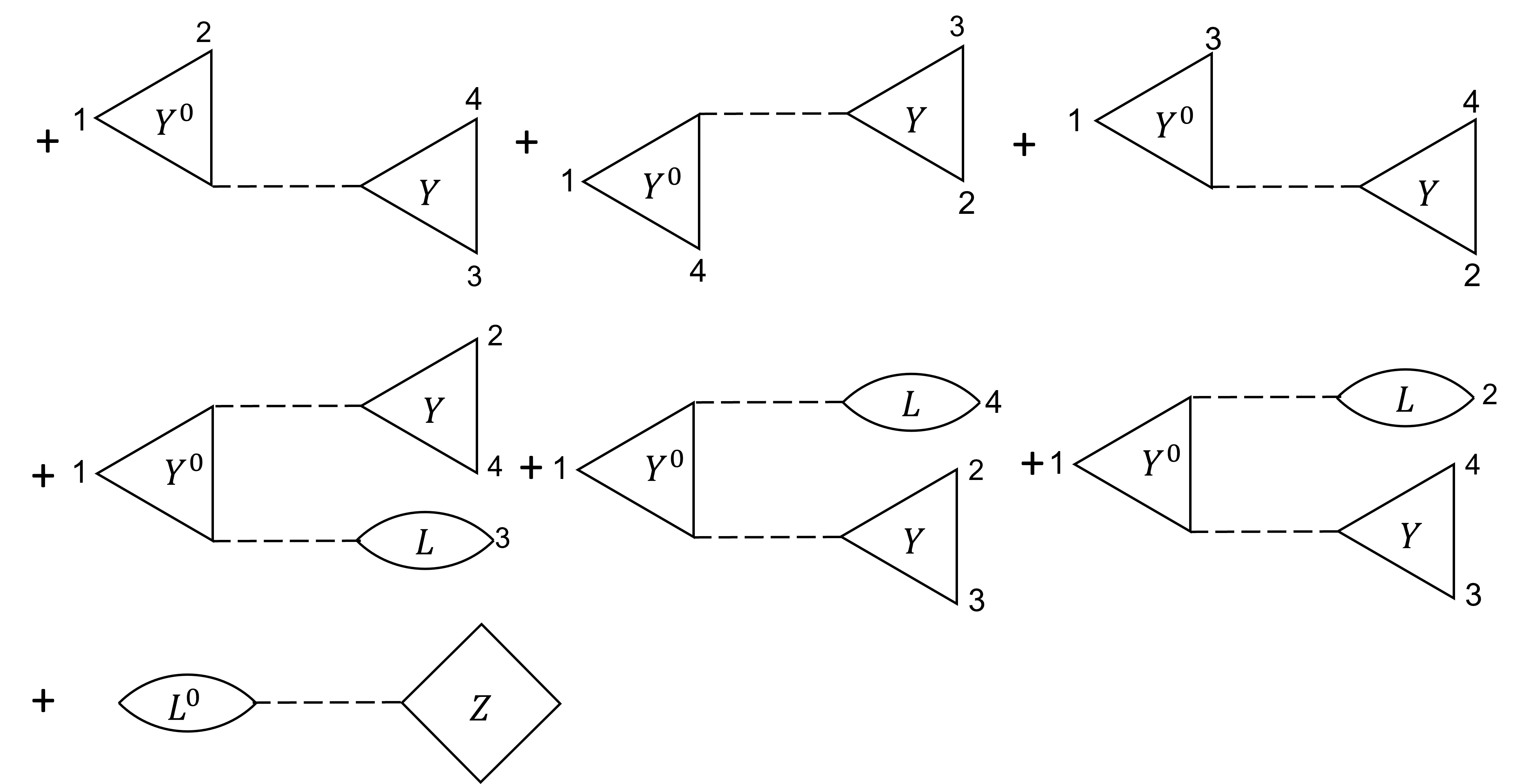}
\end{center}
\caption{\label{fig_all_f}Feynman graphs of the RPA linear, quadratic and cubic response functions as needed to compute the induced density upon harmonic perturbation.}
\end{figure*}
Fig.~\ref{fig_all_f} contains a depiction of the linear, quadratic, and cubic response functions in RPA using Feynman graphs. It illustrates nicely the topological structure of the equations of motion.

\section{Nonlinear density responses in terms of polarization functions and inverse dielectric functions\label{sec_3}}

Besides the total response functions, the effective response by way of the polarization function and the dielectric function are of interest. We recall Eq.~(\ref{expansion}) with the definitions (\ref{def_L12}), (\ref{def_Y123}), and (\ref{def_Z1234}). 

It is known that the linear response might be expressed via the polarization function $\Pi$ and the generalized inverse dielectric function $K$ via the functional derivative chain rule
\begin{align}
&L(12)=-i\frac{\delta g(11)}{\delta U(22)}=-i\int d5 \frac{\delta g(11)}{\delta U^{\mathrm{eff}}(55)}
\frac{\delta U^{\mathrm{eff}}(55)}{\delta U(22)}=\int d5 \Pi(15)K(25)\,,
\end{align}
where the effective potential is usually defined as $U^{\mathrm{eff}}=U+\Sigma^{\mathrm{H}}$, with $\Sigma^{\mathrm{H}}$ being the Hartree self energy.

In terms of functional derivatives, we thus have by analogy for the higher order general polarization functions 
\begin{align}
\Pi(12,1'2')=&\pm i \frac{\delta g(11')}{\delta U^{\mathrm{eff}}(2'2)}\,,\\
\Pi(123,1'2'3')=&(\pm i)^2\frac{\delta^2 g(11')}{\delta U^{\mathrm{eff}}(3'3)\delta U^{\mathrm{eff}}(2'2)}\,,\label{pi123}\\
\Pi(1234,1'2'3'4')=
(\pm i)^3&\frac{\delta^3 g(11')}{\delta U^{\mathrm{eff}}(4'4)\delta U^{\mathrm{eff}}(3'3)\delta U^{\mathrm{eff}}(2'2)}\,.
\end{align}
The higher order generalized inverse dielectric functions can be defined in a similar way
\begin{align}
K(12,1'2')=&\frac{\delta U^{\mathrm{eff}}(22')}{\delta U(1'1)}\,,\\
K(123,1'2'3')=&\frac{\delta^2 U^{\mathrm{eff}}(33')}{\delta U(1'1)\delta U(2'2)}\,,\\
K(1234,1'2'3'4')=&\frac{\delta^3 U^{\mathrm{eff}}(44')}{\delta  U(1'1)\delta U(2'2)\delta U(3'3)}\,.
\end{align}
From these definitions, the connections between the higher order inverse dielectric functions and the response functions follow as
\begin{align}
K(12,1'2')=&\delta(21')\delta(2'1)+\int d3\; V(23)L(31,3^+1')\delta(22')\,,\\
K(123,1'2'3')=&\mp i\int d4\;V(24)Y(421,4^+2'1')\delta(33')\,,\\
K(1234,1'2'3'4')=& -\int d5\; V(45)Z(5321,5^+3'2'1')\delta(44')\,.
\end{align}

The overall result for the density fluctuation written using the effective quantities of the polarization functions and the inverse dielectric functions up to third order is then
\bea
\delta n(1)&=&\int d2 d5\; \Pi(15)K(25)\delta U(22)\nonumber\\
&&+\frac{1}{2}\int d2 d3 d5 d6\; \Pi(156)K(36)K(25)\delta U(22)\delta U(33)\nonumber\\
&&+\frac{1}{2}\int d2 d3 d5\; \Pi(15)K(325)\delta U(22)\delta U(33)\nonumber\\
&&+\frac{1}{6}\int d2 d3 d4 d5 d6 d7\; \Pi(1567)K(47)K(36)K(25)\delta U(22)\delta U(33)\delta U(44)\nonumber\\
&&+\frac{1}{6}\int d2 d3 d4 d5 d6\; \Pi(156)\nonumber\\
&&\qquad\times\Big[
K(436)K(25)+K(425)K(36)+K(325)K(46)
\Big]\delta U(22)\delta U(33)\delta U(44)\nonumber\\
&&+\frac{1}{6}\int d2 d3 d4 d5\; \Pi(15)K(4325))\delta U(22)\delta U(33)\delta U(44)\,.
\label{dnlpi}
\eea
The first term contains the linear response. The second and third summand contain the quadratic response in terms of the linear quantities, the quadratic polarization function $\Pi(156)$ and the quadratic inverse dielectric function $K(325)$. The last three adders show the contributions to the cubic response in terms of the linear, quadratic and cubic quantities, in particular the cubic polarization $\Pi(1567)$ and cubic inverse dielectric function $K(4325)$. 

The definitions of the total response functions in terms of the effective quantities are thus
\begin{align}
L(12,1'2')=&\int d5d6\; \Pi(15,1'6)K(26,2'5)\,,\\
Y(123,1'2'3')=&
\int d5 d6d7d8\; \Pi(158,1'67)K(37,3'8)K(26,2'5)\nonumber\\
&\pm i\int d5d6\;\Pi(15,1'6)K(326,3'2'5)\,,\\
Z(1234,1'2'3'4')=&\int d5 d6 d7 d8 d\bar{5}d\bar{6} \; \Pi(158\bar{6},1'67\bar{5})K(4\bar{5},4'\bar{6})K(37,3'8)K(26,2'5)\nonumber\\
\pm i\int d5 d6d7d8\;&\Pi(158,1'67)\Big[K(437,4'3'8)K(26,2'5)
+K(426,4'2'5)K(37,3'8)\nonumber\\
&\qquad\qquad\qquad+K(326,3'2'5)K(47,4'8)\Big]\nonumber\\
&+\int d5d6\;\Pi(15,1'6)K(4326,4'3'2'5)\,.
\end{align}
The quantities needed for the density response in Eq.~(\ref{dnlpi}) are special cases of these general definitions, i.e., $\Pi(156)=\Pi(156,1'5'6')\delta(11')\delta(55')\delta(66')$.

Equations of motion can naturally be derived for all effective quantities alike. Well known is the equation for the linear polarization function which can be derived in a similar fashion to the equation for the linear response function~\cite{kremp_book}
\begin{equation}
\Pi(12,1'2')=\pm i g(12')g(21')\\
\pm i\int d3 d3' 4 4'\; g(13)g(3'1')\bar{\theta}(34,3'4')\Pi(4'2,42')\,,
\end{equation}
with the derivative of the screened self energy $\bar{\theta}(34,3'4')=\delta\bar{\Sigma}(33')/\delta g(4'4)$. In general, the equations for the effective higher order polarization functions have an identical mathematical structure as the equations for $L,\, Y$, and $Z$; only the local field corrections need to be replaced by the effective LFCs $\theta \rightarrow \bar{\theta}$. The second and third order LFCs are identical anyway.

We still give equations of motion for the total response functions in terms of effective response functions (without dielectric functions). The linear response function is well known to be determined by the linear polarization function~\cite{kremp_book}
\begin{align}
L(12,1'2')=&\Pi(12,1'2')+\int d3d5\;\Pi(15,1'5)V(53)L(31,3^+1')\,.
\end{align}
For the specific case of the density response $L(12)=L(12,1'2')\delta(11')\delta(22')$, this becomes a useful equation with convolution structure. For the quadratic response function, we obtain
\begin{align}
Y(123,1'2'3')=&\Pi(123,1'2'3')+\int d4d5\;\Pi(143,1'43')V(45)L(52,5^+2')\nonumber\\
&+\int d4d5\;\Pi(124,1'2'4)V(45)L(53,5^+3')\nonumber\\
&+\int d4d5d6d7\;\Pi(164,1'64)V(45)L(53,5^+3')V(67)L(72,7^+2')\nonumber\\
&+\int d4d5\;\Pi(14,1'4)V(45)Y(523,5^+2'3')\,.
\end{align}
As we can see, the total quadratic response is a sum of the pure quadratic polarization, terms where linear response functions are coupled to the quadratic polarization and a term multiplying the linear polarization with the quadratic response. Again, this will feature a convolution structure for the actual case of the density response. Finally, the cubic response function is
\begin{align}
Z(1234,1'2'3'4')=&\Pi(1234,1'2'3'4')
+\int d5d6\;\Pi(1534,1'53'4')V(56)L(62,6^+2')\nonumber\\
&+\int d5d6\;\Pi(1254,1'2'54')V(56)L(63,6^+3')\nonumber\\
&+\int d5d6\;\Pi(1235,1'2'3'5)V(56)L(64,6^+4')\nonumber\\
&+\int d5d6d7d8\;\Pi(1574,1'574')V(56)L(62,6^+2')V(78)L(83,8^+3')\nonumber\\
&+\int d5d6d7d8\;\Pi(1537,1'53'7)V(56)L(62,6^+2')V(78)L(84,8^+4')\nonumber\\
&+\int d5d6d7d8\;\Pi(1257,1'257)V(56)L(63,6^+3')V(78)L(84,8^+4')\nonumber\\
&+\int d5d6d7d8d\bar{5}d\bar{6}\;\Pi(157\bar{5},1'57\bar{5})V(56)L(62,6^+2')V(78)L(83,8^+3')V(\bar{5}\bar{6})L(\bar{6}4,\bar{6}^+4')\nonumber\\
&+\int d5d6\;\Pi(125,1'2'5)V(56)Y(634,6^+3'4')\nonumber\\
&+\int d5d6\;\Pi(153,1'53')V(56)Y(624,6^+2'4')\nonumber\\
&+\int d5d6\;\Pi(154,1'54')V(56)Y(623,6^+2'3')\nonumber\\
&+\int d5d6d7d8\;\Pi(175,1'75)V(56)Y(634,6^+3'4')V(78)L(82,8^+2')\nonumber\\
&+\int d5d6d7d8\;\Pi(157,1'57)V(56)Y(624,6^+2'4')V(78)L(83,8^+3')\nonumber\\
&+\int d5d6d7d8\;\Pi(157,1'57)V(56)Y(623,6^+2'3')V(78)L(84,8^+4')\nonumber\\
&+\int d5d6\;\Pi(15,1'5)V(56)Z(6234,6^+2'3'4')\;.
\end{align}
The overall mathematical structure is quite familiar by now. We 'dress' polarization functions using the linear and quadratic total response functions to calculate the cubic response function.

\section{Connection of higher order Green's functions, nonlinear density response functions, and higher order structure factors\label{sec_4}}

It is well known that the linear response function is connected to the two-particle Green's function via~\cite{green_book}
\beq
g_2(12,1'2')= -iL(12,1'2') +  g(11')g(22')\,.
\label{g2l}
\eeq
The quadratic response function is given by Eq.~(\ref{def_Y123123}), whereas the three-particle Green's function $g_3$ follows~\cite{green_book}
\beq
g_3(123,1'2'3')=\pm\frac{\delta g_2(12,1'2')}{\delta U(3'3)}+g_2(12,1'2')g(33')\,.
\eeq
These lead to the connection between $g_3$ and $Y$
\begin{align}
g_3(123,1'2'3')&=-Y(123,1'2'3')-iL(12,1'2')g(33')-iL(13,1'3')g(22')\nonumber\\
&\quad\,-iL(23,2'3')g(11')+g(11')g(22')g(33')\,.
\end{align}
Similarly, the cubic response function is defined by Eq.~(\ref{def_Z12341234}), whereas the four-particle Green's function $g_4$ follows
\begin{equation}
g_4(1234,1'2'3'4')=\pm\frac{\delta g_3(123,1'2'3')}{\delta U(4'4)}
+g_3(123,1'2'3')g(44')\,,
\end{equation}
Therefore, the connection between $g_4$ and $Z$ reads as
\bea
g_4(1234,1'2'3'4')&=&iZ(1234,1'2'3'4')\nonumber\\
&&-Y(123,1'2'3')g(44')-Y(124,1'2'4')g(33')\nonumber\\
&&-Y(134,1'3'4')g(22')-Y(234,2'3'4')g(11')\nonumber\\
&&-L(12,1'2')L(34,3'4')-L(13,1'3')L(24,2'4')\nonumber\\
&&-L(14,1'4')L(23,2'3')\nonumber\\
&&-iL(12,1'2')g(33')g(44')-iL(13,1'3')g(22')g(44')\nonumber\\
&&-iL(14,1'4')g(22')g(33')-iL(23,2'3')g(11')g(44')\nonumber\\
&&-iL(24,2'4')g(11')g(33')-iL(34,3'4')g(11')g(22')\nonumber\\
&&+g(11')g(22')g(33')g(44')
\eea
Such relations might be useful for different considerations either in the particle-particle channel or in the particle-hole channel.

\subsection{Dynamic structure factor and density fluctuations}

There is naturally an inherent connection between the density response functions, the Green's functions, and the dynamic structure factors of all orders~\cite{PhysRevE.54.3518,Kalman}. The usual dynamic structure factor, being the Fourier transform of the correlation of two density fluctuations, is given by~\cite{kremp_book}
\beq
S(\qv,\omega)=\frac{1}{2\pi N}\int\limits_{-\infty}^{\infty}d t\;
\langle \delta\rho(\qv,t)\delta\rho(-\qv,0)\rangle e^{i\omega t}\,.
\label{usualsqw}
\eeq
Defining the intermediate scattering function 
\beq
F(\qv,t)=\frac{1}{N}\langle \delta\rho(\qv,t)\delta\rho(-\qv,0)\rangle\,,
\eeq
we can rewrite this as
\beq
S(\qv,\omega)=\frac{1}{2\pi}\int\limits_{-\infty}^{\infty}dt\;
F(\qv,t) e^{i\omega t}\,.
\eeq
The inverse Fourier transform is simply~\cite{dornheim_prl_18}
\beq
F(\qv,t)=\int\limits_{-\infty}^{\infty}d\omega\;
S(\qv,\omega) e^{-i\omega t}\,.
\eeq
Considering the intermediate scattering function on the imaginary time axis within the interval $[0,-i\beta]$, we directly obtain~\cite{dornheim_prl_18}
\beq
F(\qv,\tau)=\int\limits_{-\infty}^{\infty}d\omega\;
S(\qv,\omega)e^{-\omega\tau}\,,
\label{inverseprob}
\eeq
which is a two-sided Laplace transform and gives the connection between the imaginary time correlation function, which is a standard output from path integral Monte Carlo, and the dynamic structure factor. The inversion of Eq.~(\ref{inverseprob}) is often referred to as \textit{analytic continuation} in the literature~\cite{Jarrell1996} and constitutes a formidable and, in fact, ill-posed problem. Still, due to its high relevance for the modeling of dynamic properties of non-ideal quantum many-body systems from first principles, a host of methods has been explored over the last decades, e.g., Refs.~\cite{dynamic_folgepaper,dornheim_dynamic,Vitali_PRB_2010,Fournier_PRL_2020,Filinov_PRA_2012}.Indeed, $F(\mathbf{q},\tau)$ gives one straightforward access to a host of physical properties~\cite{Dornheim_MRE_2023,Tolias_JCP_2024}, such as all existing frequency moments of $S(\mathbf{q},\omega)$~\cite{Dornheim_moments_2023}; this is in contrast to the well-known sum-rules, which are limited to odd powers in the frequency~\cite{tkachenko_book}. Finally, we note that $F(\mathbf{q},\tau)$ has recently emerged as a valuable tool for the model-free interpretation of XRTS experiments with warm dense matter and beyond~\cite{Dornheim_response_review_2023,Dornheim_T_2022,Dornheim_T_follow_up,Dornheim_SciRep_2024,Schoerner_PRE_2023}. The static structure factor may be obtained using~\cite{kremp_book}
\beq
S(\qv)=F(\qv,0)=\frac{1}{N}\langle\delta\rho(\qv,0)\delta\rho(-\qv,0)\rangle\,,
\eeq
which holds for $F$ defined in real time ($t=0$) or imaginary time ($\tau=0$).

Similarly, the correlation between three density fluctuations, the quadratic response function, may be defined as~\cite{Dornheim_imaginary_nonlinear_2021}
\beq
Y^>(123)=\langle\delta\rho(1)\delta\rho(2)\delta\rho(3)\rangle\,.
\label{yfluccor}
\eeq
Thus, the quadratic intermediate scattering factor is~\cite{PhysRevE.54.3518}
\beq
F(\qv_1,t_1;\qv_2,t_2;\qv_3,t_3)=\frac{1}{N(N-1)}\langle\delta\rho(\qv_1,t_1)\delta\rho(\qv_2,t_2)\delta\rho(\qv_3,t_3)\rangle\,.
\eeq
For homogeneous systems in equilibrium, only the relative variables are important and we can write
\begin{equation}
F(\qv_2+\qv_3,0;-\qv_2,-t_2;-\qv_3,-t_3)=\frac{1}{N(N-1)}
\langle\delta\rho(\qv_2+\qv_3,0)\delta\rho(-\qv_2,-t_2)\delta\rho(-\qv_3,-t_3)\rangle\,.
\end{equation}
Therefore, the quadratic dynamic structure factor is given by
\begin{multline}
S(\qv_2,\omega_2;\qv_3,\omega_3)=\frac{1}{N(N-1)}\frac{1}{(2\pi)^2}\int\limits_{-\infty}^{\infty}dt_2dt_3 e^{i\omega_2t_2+i\omega_3t_3}\\
\times\langle\delta\rho(\qv_2+\qv_3,0)\delta\rho(-\qv_2,-t_2)\delta\rho(-\qv_3,-t_3)\rangle
\end{multline}
The inverse transformation in imaginary time is then
\begin{equation}
F(\qv_2+\qv_3,0;-\qv_2,\tau_2;-\qv_3,\tau_3)=\\
\int\limits_{-\infty}^{\infty}d\omega_2d\omega_3\,
S(\qv_2,\omega_2;\qv_3,\omega_3)
e^{-\omega_2\tau_2-\omega_3\tau_3}
\end{equation}
The quadratic static structure factor follows as 
\begin{align}
S(\qv_2,\qv_3)&=\int\limits_{-\infty}^{\infty}d\omega_2\int\limits_{-\infty}^{\infty}d\omega_3 S(\qv_2,\omega_2;\qv_3,\omega_3)\\
&=F(\qv_2+\qv_3,0;-\qv_2,0;-\qv_3,0)\\
&=\frac{1}{N(N-1)}\langle\delta\rho(\qv_2+\qv_3,0)\delta\rho(-\qv_2,0)\delta\rho(-\qv_3,0)\rangle\,.
\end{align}

Finally, for the cubic dynamic structure factor, we have 
\begin{multline}
S(\qv_1,\omega_1;\qv_2,\omega_2;\qv_3,\omega_3)=
\frac{1}{N(N-1)(N-2)}\frac{1}{(2\pi)^3}\int\limits_{-\infty}^{\infty}dt_1dt_2dt_3\,
e^{i\omega_1t_1+i\omega_2t_2+i\omega_3t_3}\\
\times\langle\delta\rho(\qv_1+\qv_2+\qv_3,0)\delta\rho(-\qv_1,-t_1)\delta\rho(-\qv_2,-t_2)\delta\rho(-\qv_3,-t_3)\rangle
\end{multline}
The Laplace transform connection between the cubic intermediate scattering factor in imaginary time and the cubic dynamic structure factor is
\begin{multline}
F(\qv_1+\qv_2+\qv_3,0;-\qv_1,\tau_1;-\qv_2,\tau_2;-\qv_3,\tau_3)=\\
\int\limits_{-\infty}^{\infty}d\omega_1d\omega_2d\omega_3\;
S(\qv_1,\omega_1;\qv_2,\omega_2;\qv_3,\omega_3)
e^{-\omega_1\tau_1-\omega_2\tau_2-\omega_3\tau_3}
\end{multline}
and the cubic static structure factor is
\begin{equation}
S(\qv_1,\qv_2,\qv_3)=
\frac{1}{N(N-1)(N-2)}\langle\delta\rho(\qv_1+\qv_2+\qv_3,0)\delta\rho(-\qv_1,0)\delta\rho(-\qv_2,0)\delta\rho(-\qv_3,0)\rangle\,.
\end{equation}
There is an often used connection between Eq.~(\ref{usualsqw}) and the intensity of the x-ray Thomson scattering signal~\cite{siegfried_review,falk_wdm}. Whether higher order structure factors can be experimentally verified using similar methods is a subject of ongoing research~\cite{Sitenko, Fuchs_2015}.

\subsection{Martin-Schwinger hierarchy}

The first three equations of the Martin-Schwinger hierarchy for the s-particle Green's functions are easily derived~\cite{green_book}.
These equations can be given an integro-differential form or can be translated into integral equations analogous to the Dyson equation for the one-particle Green's function as shown here:
\bea
g(11')&=&g_0(11')\pm i\int d2 d3\; g_0(12)V(23)g_2(23,1'3^+)\,,\\
g_2(12,1'2')&=&g(11')g(22')\pm g(12')g(21')
-\int d3 d4\; g(13)\Sigma(34)g_2(42,1'2')\nonumber\\
&&\pm i\int d3 d4 g(13)V(34)g_3(324,1'2'4^+)\,,\\
g_3(123,1'2'3')&=&g(11')g_2(23,2'3')\pm g(12')g_2(23,1'3') + g(13')g_2(23,1'2')\nonumber\\
&&-\int d4 d5 \; g(14)\Sigma(45)g_3(523,1'2'3')
\nonumber\\
&&\pm i \int d4 d5\; g(14)V(45)g_4(4235,1'2'3'5^+)\,.
\eea
These equations nicely illustrate the connection between higher order correlation functions. This form of the hierarchy as presented here has the advantage that no LFC obscures the view of the actual built up of correlations and the connections of an $s$-particle correlation to $(s-1)\ldots (s-2)\ldots 1$ complexes as well as to an $s+1$ correlation. 

We concentrate now on the equation for $g_2$ and insert the  equation for $g_3$ into it to make fourth order correlations explicitly visible
\bea
g_2(12,1'2')&=&g(11')g(22')\pm g(12')g(21')
-\int d3 d4\; g(13)\Sigma(34)g_2(42,1'2')\nonumber\\
&&\pm i \int d3 d4\; g(13)V(34)\Big\{g(31')g_2(24,2'4^+)\pm g(32')g_2(24,1'4^+)\nonumber\\
&&\qquad\qquad\qquad\qquad\qquad+g(34^+)g_2(24,1'2') \Big\}\nonumber\\
&&\mp i\int d3 d4 d5 d6 \; g(13)V(34)g(35)\Sigma(56)g_3(624,1'2'4^+)\nonumber\\
&&-\int d3 d4 d5 d6 \; g(13)V(34)g(35)V(56)g_4(5246,1'2'4^+6^+)\,.
\eea
Therefore, there is a similar expression for the linear response function (taking into account only the quadratic response function)
\bea
-iL(12,1'2')&=&\pm g(12')g(21')-\int d3d4\; g(13)\Sigma(34)\left[ iL(42,1'2')+g(41')g(22')\right]\nonumber\\
&&\pm i \int d3 d4\;g(13)V(34)\Big[
-Y(324,1'2'4^+)-iL(32,1'2')g(44^+)\\
&&-iL(34,1'4^+)g(22')
-iL(24,2'4^+)g(31')+g(31')g(22')g(44^+)\Big]\,.\nonumber
\eea
For the special form of $L$ that is needed for the dynamic structure factor, we obtain
\bea
-iL(12)&=&\pm g(12)g(21)-\int d3d4\; g(13)\Sigma(34)\left[ iL(42,12)+g(41)g(22)\right]\nonumber\\
&&\pm i \int d3 d4\;g(13)V(34)\Big[
-Y(324,124^+)-iL(32,12)g(44^+)\nonumber\\
&& -iL(34,14^+)g(22)
 -iL(24)g(31)+g(31)g(22)g(44^+)\Big]\,.
\eea
This describes exactly how the correlations of three different density fluctuations influence the dynamic structure and might possibly hold a key of how to measure these correlations.

\subsection{Martin-Schwinger hierarchy for Coulomb systems}

A different path to explicitly account for quadratic density responses in the dynamic structure factor, in particular for Coulomb systems featuring long-range correlations, is given by considering the built up of fluctuations from the polarization function
\begin{align}
L(12,1'2')&=\Pi(12,1'2')
+\int d3 d4\; \Pi(13,1'3+)V(34)L(42,4^+2')\,.
\end{align}
Here, we need an equation of motion for the polarization function according to
\bea
\Pi(12,1'2')&=&\pm i g(12')g(21')+\int d3d4\; g(14)\bar{\Sigma}(43)\Pi(32,1'2')\nonumber\\
&&-\int d3d4d5\;g(14)V(45)\left\{K(53)\Pi(432,1'3^+2')+K(52'3,523)\Pi(43,1'3^+)\right\}\nonumber\\
&&-\int d3 d4 d5 d7 d8\;g(14)V(45)\Pi(82,8^+2')V(78)K(573)\Pi(43,1'3^+)\,.
\eea
Alternatively, one might write the polarization function not in terms of higher order dielectric functions and higher order polarization functions but prefer to use the higher order response functions
\bea
\Pi(12,1'2')&=&\pm i g(12')g(21')+\int d3d4\; g(14)\bar{\Sigma}(43)\Pi(32,1'2')\\
&&\pm i\int d3 d4\; g(14)V(43)Y(432,1'3^+2')\nonumber\\
&&\pm i \int d3 d4 d5 d8 \, g(14)V(43)Y(435,1'3^+5) V(58)\Pi(82,8^+2')\,.\nonumber
\eea
In the specification for the dynamic structure factor and the density fluctuations, this becomes
\beq
L(12)=\Pi(12)+\int d3 d4\; \Pi(13)V(34)L(42)\,,
\eeq
where correlations and exchange enter in the polarization functions as in
\bea
\Pi(12)&=&\pm i g(12)g(21)+\int d3d4\; g(14)\bar{\Sigma}(43)\Pi(32,12)\nonumber\\
&&-\int d3d4d5\;g(14)V(45)\left\{K(53)\Pi(432,13^+2)+K(523)\Pi(43,13)\right\}\nonumber\\
&&-\int d3 d4 d5 d7 d8\;g(14)V(45)\Pi(82)V(78)K(573)\Pi(43,13^+)\,.
\eea
or as in
\bea
\Pi(12)&=&\pm i g(12)g(21)+\int d3d4\; g(14)\bar{\Sigma}(43)\Pi(32,12)\\
&&\pm i\int d3 d4\; g(14)V(43)Y(432,13^+2)\nonumber\\
&&\pm i \int d3 d4 d5 d8 \, g(14)V(43)Y(435,13^+5) V(58)\Pi(82)\,.\nonumber
\eea
Naturally, the use of, e.g., quadratic response functions in the equation for the linear response function, i.e., the dynamic structure factor, is equivalent to the use of LFCs. In fact, the lowest order approximation of the quadratic response function will give the vertex correction term for the polarization function. Nevertheless, the full frequency and momentum resolved quadratic (or higher order) correlation function should give rise to a number of nonlinear features.

\section{Summary \& Outlook}

We have given in detail derivations of the equations of motion for the linear, quadratic, and cubic density response functions. For each of these, the ídeal expressions, the RPA approximation and possible insertions of LFCs were discussed. The structure of the most general form of the equations of motion was given and analyzed.

In addition, the concepts of the polarization function and the inverse dielectric function, which are important for long-range Coulomb systems, were generalized to higher order to rewrite the response quantities. In particular, we derived equations of motion for the linear, quadratic, and cubic polarization functions.

Furthermore, the connections between s-particle Green's functions and higher order response functions were made explicit. In the same manner, we showed how, based on higher order response functions, higher order structure factors are defined. These are closely connected to imaginary time correlation functions which are routinely available from PIMC simulations of finite temperature quantum systems. Similarly to the Martin-Schwinger-hierarchy of Green's functions, we established a hierarchy for the response functions where higher order response functions take the place of the local field corrections (derivatives of the self energy).

These expressions are very useful in systematic theories of the density fluctuations (damping, higher harmonics generation, mode coupling) or the stopping power. We expect such formulas to feature heavily in modern theories for the dynamic structure factor and the measured inelastic x-ray scattering signal, since state-of-the-art optical or x-ray drivers are able to precisely and significantly disturb the system of interest (solid, liquid, plasma, warm dense matter) as well as diagnose it such that second or third order nonlinear effects cannot only be excited specifically but also detected~\cite{Dornheim_PRL_2020}.

\section*{Data availability statement}
No datasets were generated or analysed during the current study.

\section*{Acknowledgments}
This work was partially supported by the Center for Advanced Systems Understanding (CASUS), financed by Germany’s Federal Ministry of Education and Research (BMBF) and the Saxon state government out of the State budget approved by the Saxon State Parliament.
This work has received funding from the European Union's Just Transition Fund (JTF) within the project \emph{R\"ontgenlaser-Optimierung der Laserfusion} (ROLF), contract number 5086999001, co-financed by the Saxon state government out of the State budget approved by the Saxon State Parliament. This work has received funding from the European Research Council (ERC) under the European Union’s Horizon 2022 research and innovation programme
(Grant agreement No. 101076233, "PREXTREME"). Views and opinions expressed are however those of the authors only and do not necessarily reflect those of the European Union or the European Research Council Executive Agency. Neither the European Union nor the granting authority can be held responsible for them.


\end{document}